\documentclass[sigconf,screen]{acmart}
 
\usepackage{booktabs} 

\copyrightyear{2018} 
\acmYear{2018} 
\setcopyright{acmlicensed}
\acmConference[ESEC/FSE '18]{Proceedings of the 26th ACM Joint European Software Engineering Conference and Symposium on the Foundations of Software Engineering}{November 4--9, 2018}{Lake Buena Vista, FL, USA}
\acmBooktitle{Proceedings of the 26th ACM Joint European Software Engineering Conference and Symposium on the Foundations of Software Engineering (ESEC/FSE '18), November 4--9, 2018, Lake Buena Vista, FL, USA}
\acmPrice{15.00}
\acmDOI{10.1145/3236024.3236061}
\acmISBN{978-1-4503-5573-5/18/11}

\graphicspath{{figures/}}
\DeclareGraphicsExtensions{.pdf,.png}

\usepackage{enumitem}
\usepackage{calc}

\usepackage{multirow} 
\usepackage{bm} 

\usepackage[framemethod=tikz]{mdframed}
\newmdenv[
  innerleftmargin=7pt,
  innerrightmargin=7pt,
  tikzsetting={draw=black,dashed,line width=0.5pt,dash pattern = on 4pt off 2pt},
  linecolor=white,
  backgroundcolor=white
]{dashedbox}
\newmdenv[
  innerleftmargin=7pt,
  innerrightmargin=7pt,
  tikzsetting={draw=black, line width=0.5pt},
  linecolor=black,
  backgroundcolor=white
]{normalbox}
\newmdenv[
  topline=false,
  bottomline=false,
  rightline=false,
  skipabove=\topsep,
  skipbelow=\topsep,
  innertopmargin=0pt,
  innerbottommargin=0pt,
  innerleftmargin=7pt,
  innerrightmargin=0pt,
  tikzsetting={draw=black, line width=3pt},
  linecolor=black,
  backgroundcolor=white
]{verticalline}

\usepackage{colortbl}   
\definecolor{VeryLightGray}{rgb}{0.92,0.92,0.92}

\usepackage{subfig}

\usepackage{quoting}
\quotingsetup{font={small}}
\quotingsetup{leftmargin=1em}
\quotingsetup{rightmargin=1em}
\quotingsetup{vskip=1.2ex}

\begin{document}
\title{Towards a Theory of Software Development Expertise}

\author{Sebastian Baltes}
\orcid{0000-0002-2442-7522}
\affiliation{%
  \institution{University of Trier}
  \city{Trier}
  \country{Germany}
}
\email{research@sbaltes.com}

\author{Stephan Diehl}
\orcid{0000-0002-4287-7447}
\affiliation{%
  \institution{University of Trier}
  \city{Trier}
  \country{Germany}
}
\email{diehl@uni-trier.de}


\begin{abstract}
Software development includes diverse tasks such as implementing new features, analyzing requirements, and fixing bugs.
Being an expert in those tasks requires a certain set of skills, knowledge, and experience.
Several studies investigated individual aspects of software development expertise, but what is missing is a comprehensive theory.
We present a first conceptual theory of software development expertise that is grounded in data from a mixed-methods survey with 335 software developers and in literature on expertise and expert performance.
Our theory currently focuses on programming, but already provides valuable insights for researchers, developers, and employers.
The theory describes important properties of software development expertise and which factors foster or hinder its formation, including how developers' performance may decline over time.
Moreover, our quantitative results show that developers' expertise self-assessments are context-dependent and that experience is not necessarily related to expertise.
\end{abstract}

%
%



\maketitle

\section{Introduction}
\label{sec:introduction}

An \emph{expert} is, according to Merriam-Webster, someone ``having or showing special skill or knowledge because of what [s/he has] been taught or what [s/he has] experienced''~\cite{MerriamWebster.com2016b}.
K. Anders Ericsson, a famous psychologist and expertise researcher, defines \emph{expertise} as ``the characteristics, skills, and knowledge that distinguish experts from novices and less experienced people''~\cite{Ericsson2006}.
For some areas, such as playing chess, there exist representative tasks and objective criteria for identifying experts~\cite{Ericsson2006}.
In software development, however, it is more difficult to find objective measures for quantifying expert performance~\cite{MockusHerbsleb2002}. 
Bergersen et al. proposed an instrument to measure programming skill~\cite{BergersenSjobergOthers2014}, but their approach may suffer from learning effects because it is based on a fixed set of programming tasks. 
Furthermore, aside from programming, software development involves many other tasks such as requirements engineering, testing, and debugging~\cite{SingerLethbridgeOthers1997, LaTozaVenoliaOthers2006, SonnentagNiessenOthers2006}, in which a software development expert is expected to be good at.

In the past, researchers investigated certain aspects of software development expertise (SDExp) such as the influence of programming experience~\cite{SiegmundKaestnerOthers2014}, desired attributes of software engineers~\cite{LiKoOthers2015}, or the time it takes for developers to become ``fluent'' in software projects~\cite{ZhouMockus2010}.
However, there is currently no theory combining those individual aspects. 
Such a theory could help structuring existing knowledge about SDExp in a concise and precise way and hence facilitate its communication~\cite{HannaySjobergOthers2007}.
Despite many arguments in favor of developing and using theories~\cite{vandeVen1989, HerbslebMockus2003, KitchenhamPfleegerOthers2002, Ralph2018}, theory-driven research is not very common in software engineering~\cite{SjobergDybaOthers2008}. 

With this paper, we contribute a theory that describes \emph{central properties} of SDExp and \emph{important factors} influencing its \emph{formation}.
Our goal was to develop a \emph{process theory}, that is a theory intended to explain and understand ``how an entity changes and develops'' over time~\cite{Ralph2018}.
In our theory, the entities are individual software developers working on different software development tasks, with the long-term goal of becoming experts in those tasks.
This fits the definition of a \emph{teleological process theory}, where an entity ``constructs an envisioned end state, takes action to reach it and monitors the progress''~\cite{vandeVenPoole1995}.
The theory is grounded in data from a mixed-methods survey with 335 participants and in literature on expertise and expert performance.
Our expertise model is task-specific, but includes the notion of transferable knowledge and experience from related fields or tasks.
On a conceptual level, the theory focuses on factors influencing the formation of SDExp over time.
It is a first step towards our long-term goal to build a \emph{variance theory}~\cite{MarkusRobey1988, Langley1999} to be able explain and predict why and when a software developer reaches a certain level of expertise~\cite{Gregor2006, Ralph2018}. 

The theory can help researchers, software developers as well as employers.
\emph{Researchers} can use it to design studies related to expertise and expert performance, and in particular to reflect on the complex relationship between experience and expertise (see Section~\ref{sec:experience-and-expertise}), which is relevant for many self-report studies.
\emph{Software developers} can learn which properties are distinctive for experts in their field, which behaviors may lead to becoming a better software developer, and which contextual factors could affect expertise development.
If they are already ``senior'', they can learn what other developers expect from good mentors or which effects age-related performance decline may have on them.
Finally, \emph{employers} can learn what typical reasons for demotivation among their employees are, hindering developers to improve, and how they can build a work environment supporting expertise development of their staff.

\section{Research Design}
\label{sec:research-design}

\begin{figure}[t]
\centering
\includegraphics[width=1\columnwidth,  trim=0.0in 0.4in 0.0in 0.2in]{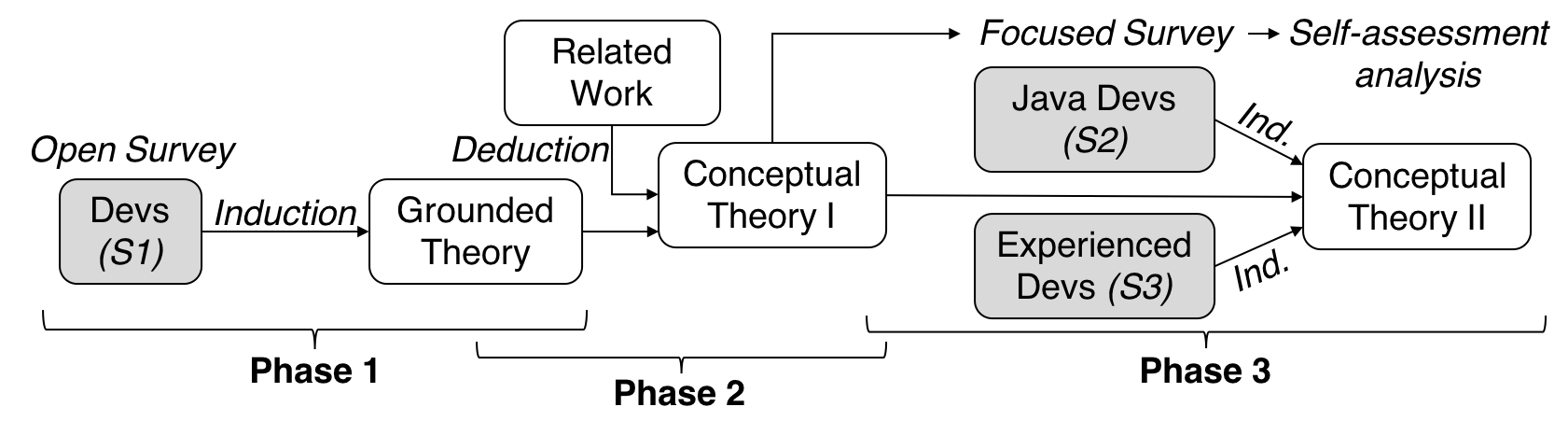} 
\caption{Research design}
\label{fig:research-design}
\end{figure}

To describe our research design, we follow Tashakkori and Teddlie's methodology~\cite{TashakkoriTeddlie1998}.
We designed a \emph{sequential mixed model study} (type VIII) with three phases (see Figure~\ref{fig:research-design}).
We started with an open online survey, which we sent out to a random sample of GitHub developers (S1) to build a preliminary grounded theory of SDExp (see Section~\ref{sec:phase1}).
In a second phase, we combined the preliminary grounded theory from the first phase with existing literature on expertise and expert performance.
The result of this combination of \emph{inductive} and \emph{deductive} methods~\cite{Gregor2006} was a preliminary conceptional theory of SDExp (see Section~\ref{sec:phase2}).
In a third phase, we designed a focused questionnaire to collect data for building a revised conceptual theory that describes certain concepts of the preliminary theory in more detail.
We sent the focused questionnaire to two additional samples of software developers (S2 and S3).
Like in the first phase, we analyzed the qualitative data from open-ended questions, this time mapping the emerging codes and categories to the preliminary conceptual theory.

To complement our qualitative analysis, we conducted a quantitative analysis investigating developers' self-assessment of programming expertise and its relation to experience (see Section~\ref{sec:experience-and-expertise}).
Please note that we planned the general research design, in particular the transitions between inductive and deductive steps~\cite{Langley1999}, before collecting the data.
We provide all questionnaires, coding schemes, and all non-confidential survey responses as supplementary material~\cite{BaltesDiehl2018}.

\section{Phase 1: Grounded Theory}
\label{sec:phase1}

\begin{table*}[t]
\centering
\scriptsize
\caption{Demographics of participants in samples $S1$-$S3$: Work time dedicated to sw. dev.; $~\text{GE}$: general experience (years), $\text{GR}_\textit{sem}$: general expertise rating (semantic differential from 1=novice to 6=expert), $\text{JE}$: Java experience (years), $\text{JR}_\textit{sem}$: Java expertise.} 
\vspace{-0.8\baselineskip}
\begin{tabular}{c|rrc|rrc|rr|rr|rr|rr|r}
\hline
\multirow{2}{*}{Sample} & \multicolumn{3}{c|}{Age} & \multicolumn{3}{c|}{Work Time (\%)} & \multicolumn{2}{c|}{$\text{GE}$ (years)} & \multicolumn{2}{c|}{$\text{JE}$ (years)} & \multicolumn{2}{c|}{$\text{GR}_\textit{sem}$ (1-6)} & \multicolumn{2}{c|}{$\text{JR}_\textit{sem}$ (1-6)} & \multirow{2}{*}{n}\\
& $M$ & $SD$ & $Mdn$ & $M$ & $SD$ & $Mdn$ & $M$ & $Mdn$ & $M$ & $Mdn$ & $M$ & $Mdn$ & $M$ & $Mdn$ & \\
\hline
$S1$ & $30.4$ & $6.4$ & $29.0$ & $70.2$ & $26.3$ & $80$ & $11.8$ & $10.0$ & $5.0$ & $3.5$ & $4.8$ & $5.0$ & $3.6$ & $4.0$ & $122$ \\
$S2$ & $31.6$ & $10.0$ & $30.0$ & $69.5$ & $26.4$ & $80$ & $12.7$ & $10.0$ & $7.6$ & $6.0$ & $4.8$ & $5.0$ & $4.4$ & $5.0$ & $127$ \\
$S3$ & $59.9$ & $4.9$ & $59.0$ & $68.2$ & $32.0$ & $80$ & $34.1$ & $35.0$ & $5.7$ & $1.5$ & $5.3$ & $5.0$ & $2.8$ & $2.0$ & $86$ \\
\hline
\end{tabular}
\label{tab:demographic-data}
\end{table*}

The goal of the first phase of our research was to build a \emph{grounded theory} (GT) of SDExp.
The GT methodology, introduced by Glaser and Strauss in 1967~\cite{GlaserStrauss1967}, is an approach to generate theory from qualitative data.
Since its introduction, different approaches evolved: Glaser's school emphasized the inductive nature of GT, while Strauss and Corbin focused on systematic strategies and verification~\cite{CorbinStrauss2008, Charmaz2014}.
The third and most recent school of GT, called \emph{constructivist} GT, tried to find a middle ground between the two diverging schools by building upon the flexibility of Glaser and Strauss's original approach, combining it with constructivist epistemology~\cite{Charmaz2014}.

All three schools rely on the process of \emph{coding} that assigns ``summative, salient, essence-capturing'' words or phrases to portions of the unstructured data~\cite{Saldana2015}.
Those codes are iteratively and continuously compared, aggregated, and structured into higher levels of abstractions, the \emph{categories} and \emph{concepts}.
This iterative process is called \emph{constant comparison}.
We followed Charmaz's constructivist approach, dividing the analysis process into three main phases: (1) \emph{initial coding}, (2) \emph{focused coding and categorizing}, and (3) \emph{theory building}.
The last step tries to draw connections between the abstract concepts that emerged from the data during the first two phases, generating a unifying theory.
An important aspect of GT is that the abstractions can always be traced back to the raw data (grounding).
In the first step, the initial coding, it is important to remain open and to stick closely to the data~\cite{Charmaz2014}.
Glaser even suggests not to do a literature review before conducting GT research~\cite{McCallin2003}, which is a rather extreme and debatable position~\cite{Thornberg2012}.
We decided to limited our literature review in the first phase to software engineering literature and postponed the integration of results from psychology literature to the second phase of our research.
The main research questions guiding this phase were:

\begin{itemize}[labelindent=0.5\parindent, labelwidth=\widthof{\textbf{RQ1:}}, label=\textbf{RQ1:}, leftmargin=*, align=parleft, parsep=0pt, partopsep=0pt, topsep=1ex, noitemsep]
\item[\textbf{RQ1:}] Which characteristics do developers assign to novices and which to experts?
\item[\textbf{RQ2:}] Which challenges do developers face in their daily work?
\end{itemize}

Our main area of interest were the characteristics developers assign to novices and experts (RQ1).
However, as software development experts are expected to master complex tasks efficiently~\cite{ZhouMockus2010}, we included a question about challenges developers face in their daily work (RQ2) to identify such tasks.

\subsection{Survey Design and Sampling}

To answer our research questions, we designed an online questionnaire, which we sent to a random sample of software developers.
Our goal was to start open-minded, thus we primarily relied on open-ended questions for data collection.
The questionnaire contained seven open-ended and four closed-ended questions related to SDExp plus seven demographic questions.
To prevent too broad and general answers, we focused on expertise in one particular programming language.
We chose Java, because at the time we designed the survey (October 2015) it was, according to various rankings, the most popular programming language~\cite{TIOBEsoftwareBV2016, Carbonnelle2016}.
We analyzed all open-ended questions separately using Charmaz's grounded theory approach, performing all three constructivist GT phases (see above) on the survey answers.
After deductively revising the resulting GT (see Section~\ref{sec:phase2}), we used theoretical sampling to collect more data on certain concepts and again performed those three GT phases, constantly comparing the new data to the data from the first iteration (see Section~\ref{sec:phase3}).
We used the closed-ended questions to describe the samples and to analyze the relation between experience and (self-assessed) expertise (see Section~\ref{sec:experience-and-expertise}).

Qualitative researchers often rely on convenience sampling for selecting their participants~\cite{Morse2007, BaltesDiehl2016}.
However, we wanted to reach a diverse sample of novices and experts, which is hard to achieve with this sampling approach.
Therefore, we drew our first sample randomly from all users who were active on both \emph{Stack Overflow} (SO) and \emph{GitHub} (GH) between January 1, 2014 and October 1, 2015. 
Both platforms are very popular among software developers and for both of them, demographic information about users is publicly available~\cite{VasilescuFilkovOthers2013}.
Another motivation for this sampling strategy was to be later able to correlate the self-assessments of developers with their activity on GH and SO. 

We derived our sampling frame from the data dumps provided by \textit{Stack Exchange} (August 18, 2015)~\citep{StackExchangeInc2015} and \textit{GHTorrent} (September 25, 2015)~\citep{Gousios2013}.
To match users on both platforms, we followed the approach of Vasilescu et al.~\cite{VasilescuFilkovOthers2013}, utilizing the MD5 hash value of users' email addresses.
For the SO users, we retrieved the email hashes from an old data dump released September 10, 2013 where this information was directly available for all users.
Further, for users who set a Gravatar URL in their profile, we extracted the email hash from there.
In the end, we were able to retrieve the email hashes for 3.8 million SO users (75\% of all users in the 2015 dataset).
In the GHTorrent data dump, the email address was available for 6.6 million GH users (69\% of all users in the dataset).
To identify active users, we checked if they contributed to a question (asked, answered, or commented) on SO and committed to a project on GH since January 1, 2014.
This resulted in a sampling frame with 71,400 unique users from which we drew a random sample of 1,000 users.
In the following, $S1$ denotes this sample.

The first iteration of the questionnaire was online from October 13, 2015 until November 11, 2015.
Of the 1,000 contacted users, 122 responded (12.2\% response rate).
Of the 122 respondents, 115 identified themselves as male, one as female and six did not provide their gender.
The majority of respondents (67.2\%) reported their main software development role to be \emph{software developer}, the second-largest group were \emph{software architects} (13.9\%).
Most participants answered from Europe (49.2\%) and North America (37.7\%).
Further demographic information can be found in Table~\ref{tab:demographic-data}.



\vspace{-0.3\baselineskip}
\subsection{Terminology}

According to Sj{\o}berg et al., the building blocks of theories are its core entities, the \emph{constructs}, the \emph{relationships} between these constructs, and the \emph{scope conditions} that delineate a theory's application area~\cite{SjobergDybaOthers2008}.
To have a consistent terminology across the paper, we use the term \emph{concepts} instead of \emph{constructs} for the central elements of the presented theories.

The \emph{scope} of all theories we built, including the GT, was to describe what constitutes SDExp and which factors influence its formation, focusing on individual developers.
In the second phase (see Section~\ref{sec:phase2}), we added a task-specific notion of expertise and then revised the resulting preliminary theory in a second inductive step (see Section~\ref{sec:phase3}) to focus on programming-related tasks.

\vspace{-0.3\baselineskip}
\subsection{Concepts}

\begin{figure}[b]
\centering
{\includegraphics[width=1\columnwidth,  trim=0.0in 0.4in 0.0in 0.2in]{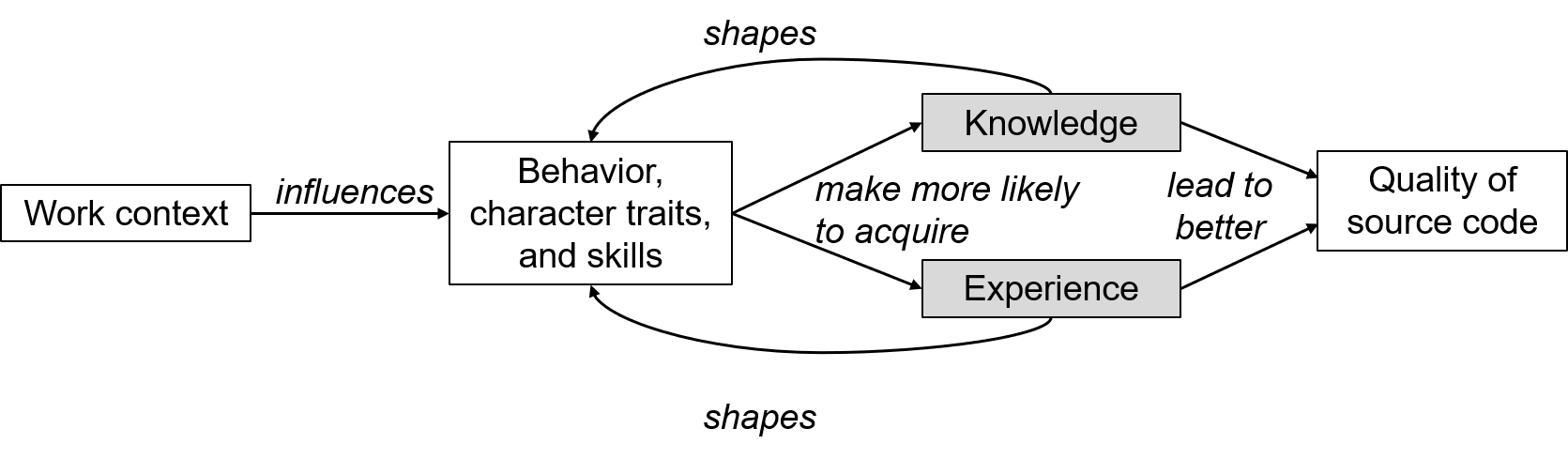} 
\caption{High-level concepts/relationships of GT (phase 1).}
\label{fig:grounded-theory}}
\end{figure}

Figure~\ref{fig:grounded-theory} shows the high-level concepts and relationships of the grounded theory that resulted from our qualitative analysis of all open-ended questions.
Most answers regarding characteristics of experts and novices (RQ1) were either related to having a certain degree of \textbf{knowledge} in different areas or a certain amount or quality of \textbf{experience}.
We marked those concepts that constitute SDExp in gray color.
The factors contributing to the formation of SDExp, and the results of having a certain degree of SDExp, have a white background.
Participants described typical \textbf{behaviors}, \textbf{character traits}, and \textbf{skills} of experts.
Many answers mentioned properties that distinguish source code written by experts from source code written by novices.
In our notion, the \textbf{quality of source code} is the result of having a certain level of knowledge and experience and thus a measure of expert performance.
When asked about challenges (RQ2), participants often named time-pressure and unrealistic demands by managers or customers.
Generally, most answers related to challenges were not technical, but referred to human factors.
In the GT, we summarized these factors as \textbf{work context}.

In the following, we present the most frequent sub-categories of the concepts mentioned above.
The concepts are in \textbf{bold} font, the (sub-)categories are in \textsc{small capitals}.
We provide a full list of all categories and subcategories as supplementary material~\cite{BaltesDiehl2018}. 

\vspace{0.25\baselineskip}
\textbf{Experience:}
Most statements that we assigned to this concept referred to the \textsc{quantity} of software development experience (e.g., in terms of years), but some also described its \textsc{quality}.
Examples for the latter include having built ``everything from small projects to enterprise projects'' or ``[experience] with many codebases''. 
In particular, participants considered \textsc{professional experience}, e.g., having ``shipped a significant amount of code to production or to a customer'' and working on \textsc{shared code} to be important factors.

\vspace{0.25\baselineskip}
\textbf{Knowledge:}
Since we specifically asked for Java, many answers were \textsc{language-specific} or referred to certain Java \textsc{frameworks}.
Experts were described as having an ``intimate knowledge of the design and philosophy of the language'' (\textsc{depth of knowledge}), which includes knowing ``the shortcomings of the language [...] or the implementation [...].''
Answers also indicated the importance of having a \textsc{broad knowledge} about algorithms, data structures, or different programming paradigms to bring ``the wisdom of [...] other philosophies into Java''.

\vspace{0.25\baselineskip}
\textbf{Quality of source code:}
Regarding the quality of source code, participants named several properties that source code of experts should possess:
It should be \textsc{well-structured} and \textsc{readable}, contain ``comments when necessary'', be ``optimized'' in terms of \textsc{performance} and sustainable in terms of \textsc{maintainability}.
One participant summarized the code that experts write as follows: ``Every one can write Java code which a machine can read and process but the key lies in writing concise and understandable code which [...] people who have never used that piece of code before [can read].''

\vspace{0.25\baselineskip}
\textbf{Behavior, character traits, and skills:}
For this concept, the most common category was \textsc{communication skills}.
Experts should be willing to ``share [their] knowledge with other developers'', but they should also know when to ``ask for help''. 
Some participants mentioned the role of experts as teachers, e.g. to train ``younger developers''. 
Another category was \textsc{(self-)reflection}, meaning reflecting on problems (``thinks before coding'') as well as on own behavior (being ``aware [of] what kind of mistakes he can make''). 
Further, participants named \textsc{problem-solving skills} and attributes that we summarized in a category named \textsc{being fast}.


\vspace{0.25\baselineskip}
\textbf{Work context:}
Many participants mentioned problems related to \textsc{people} affecting their work.
One participant put it this way: ``Computers are easy. People are hard.''
Salient were the comments about constant \textsc{time pressure}, often caused by customers or the management.
Respondents found it challenging to maintain ``quality despite pressure to just make it work''.
One participant remarked that ``sometimes non-software managers think of software like manufacturing: If 1 person works 400 parts in a day 10 should work 4000. But in software development, that analogy breaks down.''
There were also comments about team issues like ``getting a big team of developers adopt common standards of coding, designing and unit testing.''
Participants also complained about the lack of well-defined \textsc{requirements} and the importance of good \textsc{communication}:
``[...]~User's cannot communicate what they want. [...]~Project managers who talk to the users don't understand the implications by the requirements and mostly don't know enough of the business process the user lives every day. Hence, he cannot communicate the problem to the development team.''

\vspace{-0.3\baselineskip}
\subsection{Relationships}

After structuring participants' answers into concepts, categories, and sub-categories, we looked at the answers again, trying to find meaningful relationships.
The result of this process is depicted in Figure~\ref{fig:grounded-theory}.
In our notion, certain forms of \textbf{behavior}, and an individual developer's \textbf{character traits} and general \textbf{skills} make it more likely to gain the level of \textbf{knowledge} and \textbf{experience} to be considered an expert in software development, which then manifests itself in the \textbf{quality of source code} the developer creates.
However, gained knowledge and experience also affect an individual's behavior and shapes other skills.
Moreover, the \textbf{work context}, meaning, for example, the office, colleagues, customers, or the application domain of a project, influence the behavior and thus the formation of knowledge and experience.

\vspace{\baselineskip}
\begin{normalbox}
\textbf{Phase 1:}
The grounded theory describes SDExp as a combination of a certain quantity and quality of \emph{knowledge} and \emph{experience}, both general and for a particular programming language.
The \emph{work context}, \emph{behavior}, \emph{character traits}, and \emph{skills} influence the formation of expertise, which can be observed when experts write well-structured, readable, and maintainable \emph{source code}.
\end{normalbox}

\section{Phase 2: Preliminary Conceptual Theory}
\label{sec:phase2}

\begin{figure}
\centering
{\includegraphics[width=1\columnwidth,  trim=0.0in 0.4in 0.0in 0.2in]{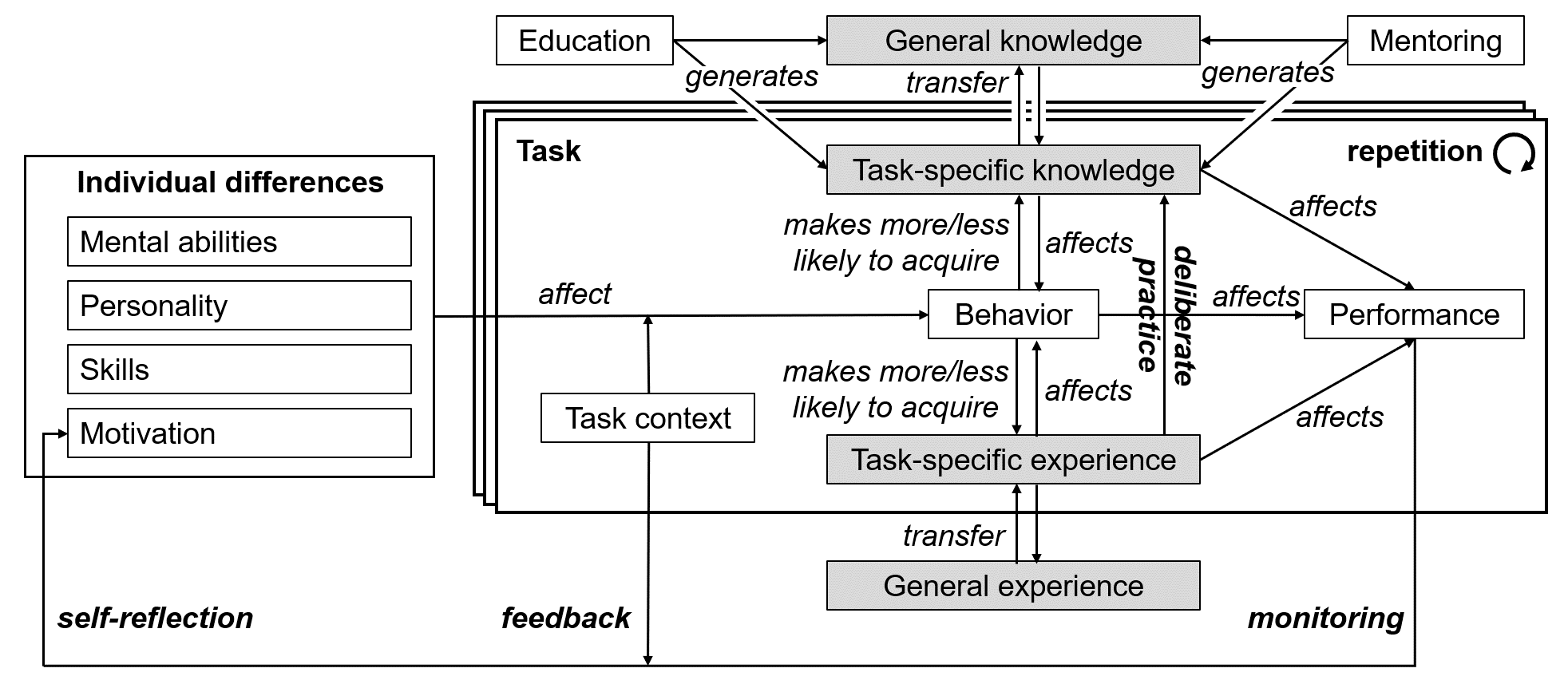} 
\caption{High-level concepts and relationships of preliminary conceptual theory (phase 2).}
\label{fig:conceptual-theory-1}}
\end{figure}

As described in our research design, the next step after \emph{inductively} deriving a preliminary GT from the responses of the participants in our first sample was to \emph{deductively} embed this GT in existing literature on expertise and expert performance.
To this end, we reviewed psychology literature.
Our main source was \emph{The Cambridge Handbook of Expertise and Expert Performance}~\cite{EricssonCharnessOthers2006} including the referenced literature.
This handbook is the first~\cite{EricssonCharnessOthers2006}, and to the best of our knowledge most comprehensive, book summarizing scientific knowledge on expertise and expert performance.
The result of this deductive step was a task-specific conceptual theory of expertise development that is compatible with the grounded theory from the first phase.
Figure~\ref{fig:conceptual-theory-1} shows our preliminary conceptual theory, which we are going to present in this section.

Generally, \emph{process theories} focus on events and try to find patterns among them, leading to a certain outcome---\emph{variance theories} describe a certain outcome as a relationship between dependent and independent variables~\cite{Langley1999}.
The process that we describe with our conceptual theory is the formation of SDExp, that is the path of an individual software development novice towards becoming an expert.
This path consists of gradual improvements with many corrections and repetitions~\cite{Ericsson2006b}, therefore we do not describe discrete steps like, for example, the 5-stage \emph{Dreyfus model} of skill acquisition (see Section~\ref{sec:self-assessments}).
Instead, we focus on the repetition of individual tasks.
In phase 3, we extended our conceptual theory with a focus on programming-related tasks (see Section~\ref{sec:phase3}), but the general structure is flexible enough to be extended towards other software development tasks as well~\cite{SingerLethbridgeOthers1997, LaTozaVenoliaOthers2006, SonnentagNiessenOthers2006}.
Even with a focus on programming expertise, the distinction between tasks is important.
For example, an excellent Java programmer is not automatically an excellent Haskell programmer.
Moreover, programming itself includes diverse tasks, such as implementing new features or fixing bugs, with a varying centrality and difficulty~\cite{ZhouMockus2010}.

\subsection{Concepts}
\label{sec:phase2-concepts}

In the following, we will describe the concepts we deductively integrated into our grounded theory using literature on expertise and expert performance.

\vspace{0.25\baselineskip}
\textbf{Individual differences and behavior:}
We split the GT concept \emph{behavior, character traits, and skills} into \emph{individual differences} and \emph{behavior}.
We modeled behavior as being relative to a certain task and as being influenced by individual differences~\cite{MotowidloBormanOthers1997} such as \emph{mental abilities}, \emph{personality}, and \emph{motivation}, which have long been considered essential for general~\cite{EricssonSmith1991, CampitelliGobet2011, HambrickMeinz2011} and programming performance~\cite{Curtis1984}.
Even if the general intelligence is not a valid predictor for attaining expert performance in a domain~\cite{Ericsson2006}, improvements are constrained by an individual's cognitive capacities~\cite{Ericsson2006b}.
Especially at the early stages of skill acquisition, general intelligence is in fact an important factor~\cite{KrampeCharness2006}.
It is also known that mental abilities start to decline at a certain age~\cite{KrampeCharness2006}

Acquiring expertise is not exclusively a cognitive matter''~\cite{Hunt2006}---developers' \emph{personality} and \emph{motivation} influence \emph{behaviors} that may or may not lead to improvements of expertise~\cite{Hunt2006, Sosniak2006}.
Generally, the term \emph{skill} is defined as ``an ability or proficiency acquired through training and practice''~\cite{AmericanPsychologicalAssociation2015}.
Thus, according to that definition, being a good software developer is also a skill.
However, in the scope of our theory, we limit the term skill to fundamental skills such as communication and social skills~\cite{AmericanPsychologicalAssociation2015}. 

\vspace{0.25\baselineskip}
\textbf{Task context:}
In the GT, we described how the \emph{work context}, including team members, managers, and customers, can influence developers' \emph{behavior}.
In the conceptual theory, we considered this context to be task-specific (e.g., communication with customers is more likely to happen during requirements analysis and communication with colleagues when refactoring an existing module).
The \emph{task context} captures all organizational, social~\cite{Mieg2006}, and technical constraints that are relevant for the task at hand.

\vspace{0.25\baselineskip}
\textbf{Knowledge and experience:}
Knowledge can be defined as a ``permanent structure of information stored in memory''~\cite{Robillard1999}.
Some researchers consider a developer's knowledge base as the most important aspect affecting their performance~\cite{Curtis1984}.
Studies with software developers suggest that ``the knowledge base of experts is highly language dependent'', but experts also have ``abstract, transferable knowledge and skills''~\cite{SonnentagNiessenOthers2006}.
We modeled this aspect in our theory by dividing the central concepts \emph{knowledge} and \emph{experience} from the GT into a task-specific and a general part.
This is a simplification of our model, because the relevance of knowledge and experience is rather a continuum than dichotomous states~\cite{Weisberg2006}.
However, Shneiderman and Mayer, who developed a behavioral model of software development, used a similar differentiation between general (``semantic'') and specific (``syntactical'') knowledge~\cite{ShneidermanMayer1979}.
General knowledge and experience does not only refer to technical aspects (e.g., low-level computer architecture) or general concepts (e.g., design patterns), but also to knowledge about and experience with successful strategies~\cite{Sonnentag1995, Sonnentag1998, KoUttl2003}.

\vspace{0.25\baselineskip}
\textbf{Performance, education, and monitoring:}
As mentioned in the introduction, it may be difficult to find objective measures for quantifying expert \emph{performance} in software development.
However, there exist many metrics and measures that can be evaluated regarding their validity and reliability for measuring expert performance.
Respondents from the first sample mentioned different characteristics of experts' source code, but also the time it takes to develop a solution.
This is related to the area of program comprehension where task correctness and response time are two important measures~\cite{DunsmoreRoper2000}.
At this point, our goal is not to treat \emph{performance} as a dependent variable that we try to explain for individual tasks, we rather consider different \emph{performance monitoring} approaches to be a means for \emph{feedback} and \emph{self-reflection}.
For our long-term goal to build a \emph{variance theory} for explaining and predicting the development of expertise, it will be more important to be able to accurately measure developers' performance.

\emph{Education} and \emph{mentoring} help building knowledge and thus contribute to the development of expertise~\cite{EricssonPrietulaOthers2007}.
Having a teacher or mentor is particularly important for \emph{deliberate practice}~\cite{EricssonPrietulaOthers2007, EricssonKrampeOthers1993}, which is a central aspect of our theory (see below).

\subsection{Relationships}
\label{sec:phase2-relationships}

The relationships in our theory are intentionally labeled with rather generic terms such as ``affects'' or ``generates'', because more research is needed to investigate them.
Nevertheless, we want to point out two central groups of relationships: \textbf{deliberate practice} and the influence of \textbf{monitoring}, \textbf{feedback}, and \textbf{self-reflection}.

\vspace{0.25\baselineskip}
\textbf{Deliberate practice:}
Having more experience with a task does not automatically lead to better performance~\cite{EricssonKrampeOthers1993}.
Research has shown that once an acceptable level of performance has been attained, additional ``common'' experience has only a negligible effect, in many domains the performance even decreases over time~\cite{FeltovichPrietulaOthers2006}.
The length of experience has been found to be only a weak correlate of job performance after the first two years~\cite{Ericsson2006b}---what matters is the \emph{quality} of the experience.
According to Ericsson et al., expert performance can be explained with ``prolonged efforts to improve performance while negotiating motivational and external constraints''~\cite{EricssonKrampeOthers1993}.
For them, \emph{deliberate practice}, meaning activities and experiences that are targeted at improving the own performance, are needed to become an expert.
For software development, Zhou and Mockus found that developers can improve their performance over time by continuously increasing the difficulty and centrality of development tasks~\cite{ZhouMockus2010}, which is in line with the concept of deliberate practice.
Traditionally, research on deliberate practice concentrated on acquired knowledge and experience to explain expert performance~\cite{EricssonSmith1991, CampitelliGobet2011, HambrickMeinz2011}.
However, later studies have shown that deliberate practice is necessary, but not sufficient, to achieve high levels of expert performance~\cite{CampitelliGobet2011}---individual differences play an important role~\cite{HambrickMeinz2011} (see above).

\vspace{0.25\baselineskip}
\textbf{Monitoring, feedback, and self-reflection:}
A central aspect of \emph{deliberate practice} is monitoring one's own \emph{performance}, and getting \emph{feedback}, for example from a teacher or coach~\cite{Ericsson2006b}.
Generally, such feedback helps individuals to \emph{monitor} their progress towards goal achievement~\cite{LockeLathamOthers1990}.
Moreover, as Tourish and Hargie note,  ``[t]he more channels of accurate and helpful feedback we have access to, the better we are likely to perform.''~\cite{TourishHargie2003b}.
In areas like chess or physics, studies have shown that experts have more accurate self-monitoring skills than novices~\cite{Chi2006}.
In our model, the feedback relation is connected to the concept \emph{task context} as we assumed that feedback for a software developer most likely comes from co-workers or supervisors.
To close the cycle, monitoring and self-reflection influence a developer's \emph{motivation} and consequently his/her \emph{behavior}.
In the revised conceptual theory (see Section~\ref{sec:phase3}), we also included mentors in this feedback cycle.

\vspace{\baselineskip}
\begin{normalbox}
\textbf{Phase 2:}
The preliminary conceptual theory builds upon the grounded theory.
Among other changes, the theory introduces a \emph{task-specific} view on expertise, separates \emph{individual differences} and \emph{behavior}, and embeds the concept of \emph{deliberate practice}, including the relationships \emph{monitoring}, \emph{feedback}, and \emph{self-reflection}.
Moreover, instead of focusing on source code, it introduces the general concept of \emph{performance} as a result of having a certain level of expertise.
\end{normalbox}

\begin{figure*}
\centering
{\includegraphics[width=\textwidth,  trim=0.1in 0.3in 0.1in 0.2in]{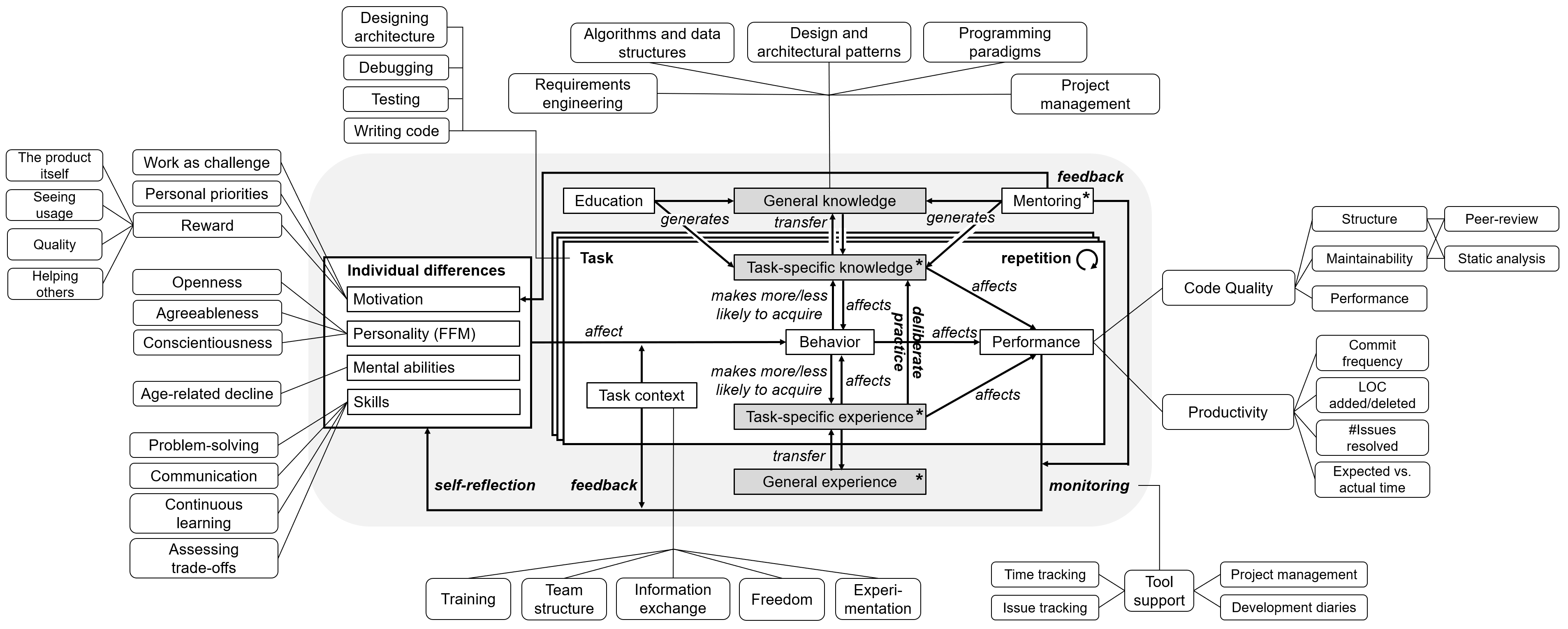} 
\caption{High-level concepts/categories of revised conceptual theory (phase 3); asterisk refers to description in the text.}
\label{fig:conceptual-theory-2}}
\end{figure*}

\section{Phase 3: Revised Conceptual Theory}
\label{sec:phase3}

The goal of the third and last phase was to validate the general design of our theory and to collect more data about certain concepts, in particular the ones related to \textbf{deliberate practice}.
Our focus was on programming-related tasks, but the theory can as well be extended and operationalized for other software development tasks in future work. 

\vspace{-0.5\baselineskip}
\subsection{Survey Design}

We revised the open questionnaire from phase 1 to focus on specific concepts, in fact most questions of the resulting focused questionnaire were directly related to individual concepts of the preliminary theory.
We then conducted \emph{theoretical sampling} to ``elaborate and refine the categories constituting [our] theory''~\cite{Charmaz2014}, surveying two additional samples of software developers.
We tried to reach active Java developers ($S2$) and very experienced developers ($S3$).
We targeted Java developers, because we wanted to compare participants' general experience and expertise with their experience and expertise in one particular programming language (see Section~\ref{sec:experience-and-expertise}).
We further targeted experienced developers, because in the first phase especially this group of participants provided well-elaborated and insightful answers.
Please note that the goal of theoretical sampling is ``conceptual and theoretical development'', not ``increasing the [...] generalizability'' of results~\cite{Charmaz2014}.

We revised and extended our two initial research questions to adjust them to our preliminary conceptual theory.
Beside asking for typical \emph{character traits} of experts (RQ1.1), we now asked in particular for traits that are supportive for becoming an expert (RQ1.2) to collect more data on factors influencing the formation of SDExp.
Due to the importance of \emph{mental abilities} in expert development and the fact that they start to decline at a certain age~\cite{KrampeCharness2006}, we asked about situations where developers' performance declined over time (RQ1.3).
Since our theory is task-specific, we also asked for \emph{tasks} that an expert should be good at (RQ1.4).
When we asked participants in S1 for challenges in their daily work (RQ2), they often referred to their \emph{work context} and in particular to people-related issues.
The work context may also influence developers' \emph{motivation}, which plays an important role in expertise development (see Section~\ref{sec:phase2-concepts}).
Thus, we changed RQ2 to focus more on those two aspects. 
Since we deductively included the concept of \emph{deliberate practice} in our theory, we added questions about \emph{monitoring} (RQ3.1) and \emph{mentoring} (RQ3.2), which are important aspects of deliberate practice.
We provide the research questions and the corresponding survey questions as supplementary material~\cite{BaltesDiehl2018}.

During the analysis of samples $S2$ and $S3$, we build upon our conceptual theory, mapping the emerging codes and categories to the existing theory.
This procedure is similar to what Salda{\~n}a calls \emph{elaborative coding}~\cite{Saldana2015}.
Figure~\ref{fig:conceptual-theory-2} depicts the high-level concepts and categories of our revised conceptual theory.
Some categories are not shown in the figure, but are described in this section.
We provide a full list of all (sub-)categories as supplementary material~\cite{BaltesDiehl2018}. 

%
%

\begin{figure*}
\centering
\includegraphics[width=0.89\textwidth,  trim=0.0in 0.2in 0.3in 0.2in]{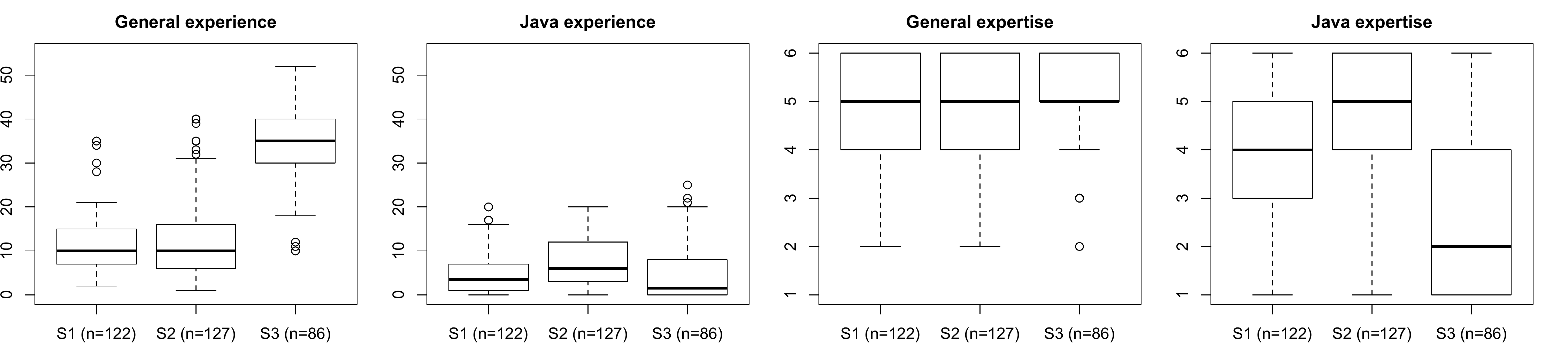} 
\caption{General/Java experience ($\text{GE}$, $\text{JE}$) and general/Java expertise rating ($\text{GR}_\textit{sem}$, $\text{JR}_\textit{sem}$) of participants in $S1$, $S2$, and $S3$.} 
\label{fig:experience-expertise}
\end{figure*}

\vspace{-0.5\baselineskip}
\subsection{Sampling}

As mentioned in the previous section, our preliminary conceptual theory guided the sampling (\emph{theoretical sampling}~\cite{Morse2007, Charmaz2014}).
Our goal was to reach active Java developers ($S2$) as well as very experienced developers ($S3$).
We retrieved the sampling frame for those samples from the \emph{Stack Exchange Data Dump}~\cite{StackExchangeInc2016} released January 1, 2016 and the \emph{GHTorrent data dump}~\cite{Gousios2013} released February 16, 2016.

For the Java sample ($S2$), we started by identifying active GH projects.
We first filtered out the projects that were not deleted, not a fork, had at least two contributing users, and had at least 10 commits.
Then, to select non-trivial Java projects, we only considered projects with at least 300 kB of Java source code (sum of file sizes of all Java files in the project).
From the resulting 22,787 Java GH projects, we created a sampling frame with all users who contributed (committed or merged a pull request) to one of the selected projects and who pushed at least 10 commits between January 1, 2015 and December 31, 2015.
From the 44,138 users who satisfied the above criteria, we contacted the ones with a public email address on their profile page ($n=1,573$).

With the third sample ($S3$), we wanted to reach very experienced users.
Therefore, we again combined data from SO and GH.
We used the age of a developer as a proxy variable for their experience.
For GH users, the age was not available, but 11\% of the users in the SO dump provided their age.
To select experienced users, we filtered all SO users with age $\geq 55$ years and $\leq 80$ years and matched them with a GH account using the hash value of their email address.
This resulted in a sample of 877 experienced users.

The focused questionnaire we used in the third phase contained nine open-ended and nine closed-ended questions, three of them only visible depending on previous answers, plus seven demographic questions.
The full questionnaire is available as supplementary material~\cite{BaltesDiehl2018}.
This iteration of the questionnaire was online from February 18, 2016 until March 3, 2016 ($S2$) and from February 19, 2016 until March 4, 2016 ($S3$).
Of the 1,573 contacted users in $S2$, 30 had an invalid email address and could not be reached.
In the end, 127 participants filled out the questionnaire (response rate  8.2\%).
Of the 877 users in $S3$, 18 had an invalid email address and 91 participants completed the questionnaire (response rate 10.6\%).
We removed five participants from $S3$ because their answers either indicated that the age information from SO was not correct or that they were not active software developers.
This lead to 86 responses available for analysis.
Overall, combining $S2$ and $S3$, we had 213 valid responses in phase 3.

In $S2$, 119 respondents identified themselves as male, three as female and five did not provide their gender ($S3$: 84/1/1).
The majority of respondents ($S2$: 64.6\%, $S3$: 61.6\%) reported their main software development role to be \emph{software developer}, the second-largest group were \emph{software architects} ($S2$: 13.4\%, $S3$: 17.5\%).
In $S2$, most participants answered from Europe (47.2\%) and North America (32.3\%), in $S3$ the order was reversed (North America (67.4\%), Europe (23.3\%)).
Further demographic information can be found in Table~\ref{tab:demographic-data}.

Comparing the demographics of the first two samples, we can see that $S1$ and $S2$ are quite similar, except for the fact that participants in $S2$ had more experience with Java ($Mdn$ $3.5$ vs. $6$ years) and rated their Java expertise to be higher ($Mdn$ $4$ vs. $5$).
This indicates that our sampling approach was successful in reaching active Java developers.
In $S3$, the values for the amount of professional work time dedicated to software development are quite similar to the other two samples.
However, the developers in this sample are much older ($M$ $59.9$ vs. $30.4$/$31.6$) and have much more general programming experience ($Mdn$ $35$ vs. $10$/$10$).
This indicates that our sampling approach for S3 was successful in reaching developers with a long general programming experience.
However, many developers in $S3$ have little Java experience ($Mdn$ $1.5$ years) and also rated their Java expertise relatively low ($Mdn$ $2$).
One reason for this could be that one quarter of the participants had a programming experience of 40 years or more ($Q_{3} = 40$) and compared to this time frame, Java is a relatively young programming language (introduced 1995).
The boxplots in Figure~\ref{fig:experience-expertise} visualize the differences in general/Java experience and expertise between the three samples.

\subsection{Concepts}

Figure~\ref{fig:conceptual-theory-2} shows the revised conceptual theory resulting from our analysis of the closed- and open-ended answers of samples $S2$ and $S3$.
In the following, we describe the most frequent (sub-)categories for the high-level concepts of our theory that emerged during the analysis and combine those qualitative results with quantitative evaluations where possible.
For each concept, we indicate when there were notable differences between the answers in $S2$ and $S3$.
Like before, we write the concepts in \textbf{bold} font and the (sub-)categories in \textsc{small capitals}.
We also provide the number of answers we assigned to each concept or category (in brackets).
We only present the most frequent categories and provide the complete coding schema as supplementary material~\cite{BaltesDiehl2018}. 

\vspace{0.25\baselineskip}
\textbf{Tasks:}
Since our SDExp model is task-specific, we asked our participants to name the three most important tasks that a software development expert should be good at.
The three most frequently mentioned tasks were \textsc{designing software architecture} (95), \textsc{writing source code} (91), and \textsc{analyzing and understanding requirements} (52).
Many participants not only mentioned the tasks, but also certain quality attributes associated with them, for example ``architecting the software in a way that allows flexibility in project requirements and future applications of the components'' and ``writing clean, correct, and understandable code''.
Other mentioned tasks include \textsc{testing} (48), \textsc{communicating} (44), \textsc{staying up-to-date} (28), and \textsc{debugging} (28).
Our theory currently focuses on tasks directly related to programming (see Figure~\ref{fig:conceptual-theory-2}), but the responses show that it is important to broaden this view in the future to include, for example, tasks related to requirements engineering (\textsc{analyzing and understanding requirements}) or the adaption of new technologies (\textsc{staying up-to-date}). 

\vspace{0.25\baselineskip}
\textbf{Experience, knowledge, and performance:}
Like in the first phase, we asked participants about general attributes of software development experts.
Aspects like having \textbf{experience} (26), a broad \textbf{general knowledge} (35) about ``paradigms [...], data structures, algorithms, computational complexity, and design patterns'', and an ``intimate'' knowledge about a certain programming language (\textbf{task-specific knowledge} (30)) were important.
In particular, knowledge about \textsc{software architecture}, including ``modularization'' and ``decomposition'', was frequently named (22).
Interestingly, 20 of the 22 answers mentioning software architecture came from the sample of active Java developers.
Also similar to the first phase, participants described properties of experts' source code such as \textsc{maintainability} (22), \textsc{clear structure} (12), or \textsc{performance} (9).
The answers from $S2$ and $S3$ supported the general structure of our theory, which we derived inductively in phase 1 and deductively in phase 2.
Thus, we will focus on new aspects and in particular on factors influencing the formation of SDExp in the following.

\vspace{0.25\baselineskip}
\textbf{Individual differences:}
We asked for specific characteristics of experts and in particular for character traits that support expertise development.
Regarding the \textbf{personality} of experts, participants often described three properties that are also present in the popular five factor personality model (FFM)~\cite{McCraeJohn1992}:
According to our participants, experts should be \textsc{open-minded} (42) and \textsc{curious} (35) (FFM: \emph{openness}), be \textsc{team players} (37) (FFM: \emph{agreeableness}), and be thorough and pay \textsc{attention to detail} (FFM: \emph{conscientiousness}).
Two other important traits were being \textsc{patient} (26) and being \textsc{self-reflected} (20).
The latter is an important connection to the concept of \textbf{deliberate practice} that we introduced in the previous phase and includes understanding one's ``weaknesses and strengths'' and ``the ability to learn from prior mistakes''.

Regarding \textbf{skills} that an expert should possess, \textsc{problem-solving} (84) was most frequently named.
Sub-categories of problem solving are \textsc{abstraction/decomposition} (30), \textsc{analytical thinking} (20), and \textsc{logical thinking} (17).
An expert can ``break a giant problem into the little pieces that can be solved to add back up to the whole''.
Examples where an analytical approach is needed include bug fixing or ``mapping the problem domain into the solution space''.
A second important skill was having the ``drive to keep learning'', which some participants described as \textsc{continuous learning} (55).
Moreover, like in the first phase, \textsc{communication skills} (42) were frequently named.
In the answers of this iteration, those skills were often mentioned together with the task of understanding and implementing \textsc{requirements} (32):
An expert should be ``a good listener during requirement gathering'', understand ``a customer's desires'', ``work out what is really needed when the client can only say what they think they want'', and should be able to ``explain what he is doing to non developers''.
According to our participants, another important skill is being able to assess \textsc{trade-offs} (19) when comparing alternative solutions.
Trade-offs can exist between ``design, maintainability, [and] performance''.
Experts should be ``able to discern the differences between early optimization and important design decisions for the long term goal'', which is closely related to the concept of \emph{technical debt} in software projects~\cite{KruchtenNordOthers2012}.


\vspace{0.25\baselineskip}
\textbf{Mentoring:}
More than half the the participants in $S2$ and $S3$ (54.3\%) had a (former) colleague or teacher whom they would describe as their mentor in becoming a better software developer.
We asked those participants to describe their mentor(s).
Six categories emerged during the initial and focused coding of participants' answers.
One category, \textsc{having time}, was only present in the answers from $S3$:
Eight experienced developers named aspects such as taking time to explain things or honoring solutions that take more time in the beginning, but save time on the long run.

Regarding the mentor's \textsc{role}, \textsc{senior developer} (15), \textsc{professor or teacher} (13) and \textsc{peer} (12) were the most common answers.
Two participants noted that their mentor was actually a \textsc{junior developer} younger than themselves.
What is important are a mentor's \textsc{character} (29), \textsc{skills} (19), his/her \textsc{experience} (16), and his/her role as a source for \textsc{feedback} (20) and as a \textsc{motivator} (19).
The most common characteristics of mentors were being \textsc{guiding} (10), \textsc{patient} (8), and \textsc{open-minded} (7).
The most important aspect of a mentor's \textsc{feedback} were comments about \textsc{code quality} (7).
What participants motivated most was when mentors posed \textsc{challenging} tasks.
In summary, we can conclude that the description of good mentors resembles the description of software development experts in general.

\vspace{0.25\baselineskip}
\textbf{Monitoring and self-reflection:}
We asked participants if they regularly monitor their software development activities.
Combining the answers from $S2$ and $S3$, 38.7\% of the 204 participants who answered that question said that they regularly monitor their activity.
We asked those participants how they usually monitor their development activity.

In both samples, the most important monitoring activity was \textsc{peer review} (16), where participants mentioned asking co-workers for feedback, doing code-review, or doing pair-programming. One participant mentioned that he tries to ``take note of how often [he] win[s] technical arguments with [his] peers''.
Participants also mentioned \textsc{time tracking} (14) tools like \textit{WakaTime} or \textit{RescueTime}, \textsc{issue tracking} (11) systems like \emph{Jira} or \emph{GitHub issues}, and \textsc{project management} (14) tools like \emph{Redmine} and \emph{Scrum story points} as sources for feedback, comparing expected to actual results (e.g., time goals or number of features to implement).
Three developers reported writing a \textsc{development diary}. 

Regarding employed metrics, participants reported using simple metrics such as the \textsc{commit frequency}, \textsc{lines of code added / deleted}, or number of \textsc{issues resolved}.
Further, they reported to use \textsc{static analysis} (18) tools such as \emph{SonarQube}, \emph{FindBugs}, and \emph{Checkstyle}, or to use \textsc{GitHub's activity overview} (10).
In this point, there was a difference between the answers in $S2$ and $S3$: GitHub's activity overview was mentioned almost exclusively by the active Java developers (9).
Three developers were doubtful regarding the usefulness of metrics.
One participant noted: ``I do not think that measuring commits [or] LOC [...] automatically is a good idea to rate performance. It will raise competition, yes---but not the one an employer would like. It will just get people to optimize whatever is measured.''
The described phenomenon is also known as \emph{Goodhart's law}~\cite{Goodhart1984, ChrystalMizen2003}.

\vspace{0.25\baselineskip}
\textbf{Motivation:}
To assess developers' motivation, we asked our participants what the most rewarding part of being a software developer is for them.
Many participants were intrinsically motivated, stating that \textsc{problem solving} (46) is their main motivation---one participant wrote that solving problems ``makes [him] feel clever, and powerful.''
Another participant compared problem solving to climbing a mountain:  ``I would equate that feeling [of getting a feature to work correctly after hours and hours of effort] to the feeling a mountain climber gets once they reach the summit of Everest.''
Many developers enjoy seeing the \textsc{result} (53) of their work.
They are particularly satisfied to see a solution which they consider to be of high \textsc{quality} (22).
Four participants mentioned refactoring as a rewarding task.
One answered: ``The initial design is fun, but what really is more rewarding is refactoring.''
Others stressed the importance of \textsc{creating something new} (19) and \textsc{helping others} (37).
Interestingly, \textsc{money} was only mentioned by six participants as a motivation for their work.



\vspace{0.25\baselineskip}
\textbf{Work context:}
To investigate the influence of the work context on expertise development, we asked what employers should do in order to facilitate a continuous development of their employees' software development skills.
We grouped the responses into four main categories: 1. \textsc{encourage learning} (70), 2. \textsc{encourage experimentation} (61), 3. \textsc{improve information exchange} (53), and 4. \textsc{grant freedom} (42).
To \textsc{encourage learning}, employers may offer in-house or pay for external \textsc{training courses} (34), pay employees to visit \textsc{conferences} (15), provide a good analog and/or digital \textsc{library} (9), and offer \textsc{monetary incentives} for self-improvement (7).
The most frequently named means to \textsc{encourage experimentation} were motivating employees to pursue \textsc{side projects} (29) and building a work environment that is open for \textsc{new ideas and technologies} (23).
To \textsc{improve information exchange} between development teams, between different departments, or even between different companies, participants proposed to \textsc{facilitate meetings} (16) such as agile retrospectives, ``Self-improvement Fridays'', ``lunch-and-learn sessions'', or ``Technical Thursday'' meetings.
Such meetings could explicitly target information exchange or skill development.
Beside dedicated meetings, the idea of developers \textsc{rotating} (15) between teams, projects, departments, or even companies is considered to foster expertise development.
To improve the information flow between developers, practices such as \textsc{mentoring} (9) or \textsc{code reviews} (8) were mentioned.
Finally, \textsc{granting freedom}, primarily in form of \textsc{less time-pressure} (18), would allow developers to invest in learning new technologies or skills.

\vspace{0.25\baselineskip}
\textbf{Performance decline:}
We asked participants if they ever observed a significant decline of their own programming performance or the performance of co-workers over time.
Combining the answers from $S2$ and $S3$, 41.5\% of the 205 participants who answered that question actually observed such a performance decline over time.
We asked those participants to describe how the decline manifested itself and to suggest possible reasons.
The main categories we assigned to those answers were: 1. different reasons for \textsc{demotivation} (34), 2. changes in the \textsc{work environment} (32), 3. \textsc{age-related decline} (13), 4. \textsc{changes in attitude} (10), and 5. \textsc{shifting towards other tasks} (7).
The most common reason for an increased \textsc{demotivation} was \textsc{non-challenging work} (8), often caused by tasks becoming routine over time.
One participant described this effect as follows: ``I perceived an increasing procrastination in me and in my colleagues, by working on the same tasks over a relatively long time (let's say, 6 months or more) without innovation and environment changes.'' 
Other reasons included not seeing a clear \textsc{vision or direction} in which the project is or should be going (7) and missing \textsc{reward} for high-quality work (6).
Regarding the \textsc{work environment}, participants named \textsc{stress} (6) due to tight deadlines or economic pressure (``the company's economic condition deteriorated'').
Moreover, bad \textsc{management} (8) or \textsc{team structure} (5) were named.
An example for bad management would be ``[h]aving a supervisor/architect who is very poor at communicating his design goals and ideas, and refuses to accept that this is the case, even when forcibly reminded.''.
\textsc{Changes in attitude} may happen due to personal issues (e.g., getting divorced) or due to shifting priorities (e.g., friends and family getting more important). 
When developers are being promoted to team leader or manager, they \textsc{shift towards other tasks}, resulting in a declining programming performance.

\textsc{Age-related decline} was described in both samples, but the more elaborate answers came from the experienced developers.
We consider the investigation of age-related performance decline in software development, together with the consequences for individual developers and the organization, to be an important area for future research.
To illustrate the effects that age-related decline may have, we provide four verbatim quotes by experienced developers:

\begin{quoting}
``In my experience (I started programming in 1962), new languages, systems, hardware became more complex and more diverse, programming became more complex. In my 50s I found it difficult to keep up with new paradigms and languages. So I turned to technical writing and eventually stopped programming.'' 
(software developer, age 72)
\end{quoting}

\begin{quoting}
``For myself, it's mostly the effects of aging on the brain. At age 66, I can't hold as much information short-term memory, for example. In general, I am more forgetful. I can compensate for a lot of that by writing simpler functions with clean interfaces. The results are still good, but my productivity is much slower than when I was younger.''
(software architect, age 66)
\end{quoting}

\begin{quoting}
``Programming ability is based on desire to achieve. In the early years, it is a sort of competition. As you age, you begin to realize that outdoing your peers isn't all that rewarding. [...] I found that I lost a significant amount of my focus as I became 40, and started using drugs such as ritalin to enhance my abilities. This is pretty common among older programmers.'' (software developer, age 60)
\end{quoting}


\begin{quoting}
``I've been in the software industry for 36 years. [...]
It seems as if for the first half or two thirds of that time I was fortunate to be involved in areas at the forefront of the technology wave [...].
For the last 10-15 years though, I have increasingly had the feeling that waves of technology were passing me by [...].
Once I do start to get involved [...] there is a huge learning curve to overcome and I labour to deliver stories as rapidly as younger colleagues who have been immersed in the relevant technology for longer.''
(software developer, age 57)
\end{quoting}

\subsection{Relationships}

The only relationships we added are related to the concept of \textbf{mentoring}.
As mentioned above, participants described mentors as an important source for \textsc{feedback} and as \textsc{motivators}.
Thus, we connected \textbf{mentoring} to the corresponding concepts \textbf{motivation} and \textbf{feedback} in the revised conceptual theory.

\vspace{\baselineskip}
\begin{normalbox}
\textbf{Phase 3:}
To refine and elaborate certain concepts of our preliminary conceptual theory, we conducted a second inductive step, collecting data from two additional samples of software developers.
We added details about \emph{individual differences} and \emph{task contexts} that foster the formation of SDExp, and further investigated concepts such as \emph{monitoring}, \emph{mentoring} and \emph{self-reflection}, which are related to \emph{deliberate practice}.
We also asked about \emph{performance decline} over time and identified \emph{age-related decline} as a problem for older software developers.
\end{normalbox}

\section{Experience and Expertise}
\label{sec:experience-and-expertise}

Since software developers' expertise is difficult to measure~\cite{MockusHerbsleb2002}, researchers often rely on proxies for this abstract concept~\cite{SiegmundKaestnerOthers2014}.
We investigated the relationship and validity of the two proxies \emph{length of experience} and \emph{self-assessed expertise} to provide guidance for researchers.


\subsection{Programming Experience vs. Expertise}

As mentioned above, we asked participants for their general and Java programming experience (years) and for a self-assessment of their general and Java expertise (semantic differential from 1=novice to 6=expert), see Table~\ref{tab:demographic-data} and Figure~\ref{fig:experience-expertise}.
To explore how experience, self-assessed expertise, and other variables are related, we employed the nonparametric \emph{Spearman's rank correlation coefficient} ($\rho$).
Our interpretation of $\rho$ is based on Hinkle et al.'s scheme~\cite{HinkleWiersmaOthers1979}: low ($0.3\leq|\rho|<0.5$), moderate ($0.5\leq|\rho|<0.7$), high ($0.7\leq|\rho|<0.9$), and very high correlation ($0.9\leq|\rho|\leq 1$).
We chose this non-parametric test because not all variables we tested had interval scaling and not all of them were normally distributed.

We highlight important correlations in the following and provide the complete correlation table as supplementary material~\cite{BaltesDiehl2018}.
For samples $S1$ and $S2$, the general experience in years ($\text{GE}$) correlates at least moderately with the self-assessed general expertise rating ($\text{GR}_\textit{sem}$) and the participants' age in years.
Interestingly, this correlation cannot be observed for the experienced developers ($S3$).
For the active Java developers ($S2$), the general experience ($\text{GE}$) and the Java experience ($\text{JE}$) have a high correlation.
The Java experience ($\text{JE}$) has a high correlation with the self-assessed Java expertise rating ($\text{JR}_\textit{sem}$) for all three samples, and a moderate correlation with the age for the active Java developers ($S2$).

From the observed correlations, we cannot draw consistent conclusions that are valid for all three samples and for both types of experience (general and Java).
Our interpretation of these results is that, depending on the background of the participants, experience in years can or cannot be a valid proxy for (self-assessed) programming expertise.
Generally, despite the fact that most researchers would probably agree with the definition of expertise as achieving ``outstanding performance''~\cite{EricssonSmith1991}, in many empirical studies programming expertise has been operationalized as years (or months) of programming experience~\cite{SiegmundKaestnerOthers2014, SonnentagNiessenOthers2006}.
Our results, which suggest that this operationalization may not be valid, is in line with studies showing that excellent software professionals have broader but not necessarily longer experience~\cite{SonnentagNiessenOthers2006, Sonnentag1995, Sonnentag1998, DiesteArandaOthers2017}.

\subsection{Validity of Expertise Self-assessments}
\label{sec:self-assessments}

In the previous subsection, we motivated that experience may not always be a valid proxy for expertise.
We were also interested in the validity of self-assessed expertise, which is, like other self-reports, context-dependent~\cite{SchwarzOyserman2001}.
The validity of self-assessed expertise is related to the concept of \textbf{self-reflection} in our conceptual theory, but has also methodological implications for software engineering research in general, because self-assessed programming expertise is often used in studies with software developers to differentiate between novices and experts~\cite{SiegmundKaestnerOthers2014}.
To analyze the influence of question context on expertise self-assessments, we asked the participants in samples $S2$ and $S3$ for a second self-assessment of their Java expertise at the end of the online survey.
At that point, we did not only provide a semantic differential scale like in the beginning of the survey ($\text{JR}_\textit{sem}$, see Table~\ref{tab:demographic-data}), but also a description of the rating scale stages based on the 5-stage \emph{Dreyfus model} of skill acquisition~\cite{DreyfusDreyfus1980} ($\text{JR}_\textit{dre}$), ranging from \emph{novice} (1) to \emph{expert} (5).
This model has been applied in various contexts, but researchers also discussed its limitations~\cite{Pena2010}.
We based our description of the Dreyfus model on a later description by Stuart Dreyfus~\cite{Dreyfus2004} and an adapted version by Andy Hunt~\cite{Hunt2008}.
We provide the description of the five stages, which we used in the focused questionnaire, as supplementary material~\cite{BaltesDiehl2018}.
The goal of this setup was to investigate if providing additional context has a significant influence on developers' self-assessment compared to a semantic differential scale without context.

When designing the initial questionnaire, we chose a 6-point scale for the expertise rating such that participants have to decide whether they consider themselves to be either on the novice (1-3) or expert (4-6) side of the scale, without the option to select a middle value~\cite{Garland1991, Moors2008}.
To be able to compare the ratings, we had to adjust $\text{JR}_\textit{sem}$ to be in range $[1,5]$ using the following function:  $adj(x)=\frac{1}{5}+\frac{4}{5}x$.
To test for significant differences between the two ratings, we applied the non-parametric two-sided \emph{Wilcoxon signed rank test}~\cite{Wilcoxon1945} and report the corresponding p-value ($p_w$).
To measure the effect size, we used \emph{Cliff's delta} ($\delta$)~\cite{Cliff1993}.
Our interpretation of $\delta$ is based on the guidelines by Kitchenham et al.~\cite{KitchenhamMadeyskiOthers2017}. 
Moreover, we report the confidence interval of $\delta$ at a 95\% confidence level ($CI_\delta$).

The Wilcoxon signed rank test indicated that $\text{JR}_\textit{dre}$ is significantly higher than $\text{JR}_\textit{sem}$ for the experienced developers in $S3$ ($p_w = 0.0009$), but the difference is not significant for the active Java developers in $S2$ ($p_w = 0.47$).
Cliff's $\delta$ shows only a negligible effect for $S2$ ($\delta = 0.08, \; CI_\delta = [-0.20, 0.04]$), but a small positive effect for $S3$ ($\delta = 0.17, \; CI_\delta = [0.004, 0.33]$), i.e., experienced developers tended to adjust their self-assessments to a higher rating after we provided context.
A possible interpretation of this result could be found in the \emph{Dunning-Kruger effect}~\cite{KrugerDunning1999}, which is one form of the \emph{illusory superiority bias}~\cite{Hoorens1993} where individuals tend to overestimate their abilities.
One result of Kruger and Dunning is that participants with a high skill-level underestimate their ability and performance relative to their peers~\cite{KrugerDunning1999}.
This may have happened in the sample with experienced developers ($S3$) when they assessed their Java expertise using the semantic differential scale.
When we provided context in form of the Dreyfus model, they adjusted their ratings to a more adequate rating, whereas the less experienced developers ($S2$) stuck to their, possibly overestimated, ratings.
We cannot conclude that the Dreyfus model in fact leads to more adequate ratings for experienced developers, because we do not have the data to assess the validity of their ratings.
However, we can conclude that the way we asked developers to assess their Java programming expertise was influenced by the context we provided.

\vspace{\baselineskip}
\begin{normalbox}
\textbf{Experience and expertise:}
Neither developers' experience measured in years nor the self-assessed programming expertise ratings yielded consistent results across all settings.
One direction for future work is to investigate and compare different expertise rating scales to provide guidance for researchers designing studies with expertise self-assessments.
\end{normalbox}



\section{Limitations and Threats to Validity}

Since we conducted mixed-methods research, we assess the limitations and threats to validity of our study in terms of the typical quantitative categories \emph{internal} and \emph{external validity}~\cite{JohnsonOnwuegbuzieOthers2007}, but we will also apply the qualitative evaluation criteria \emph{credibility}, \emph{originality}, \emph{resonance}, and \emph{usefulness}~\cite{Charmaz2014}.

\emph{Internal validity:} In our analysis of expertise self-assessments (see Section~\ref{sec:self-assessments}), we cannot rule out that a confounding factor lead to the higher self-assessments of experienced developers ($S3$).
However, although we used the same questionnaire for $S2$ and $S3$, the effect was only significant and non-negligible for $S3$.
Our goal was not to be able to quantify the effect of context on developers' self-assessment, but to show that it exists to motivate future research on this aspect.

\emph{External validity:} The main limitation affecting external validity is our focus on Java and on open source software development, in particular on GH and SO users.
Moreover, as one of three samples targeted experienced developers and only five participants identified themselves as female, our results may be biased towards experienced male developers.
Nevertheless, we are confident that our theory is also valid for other developer populations, because of the abstract nature of its core concepts and their grounding in related work.
Moreover, although we contacted open source developers, many of them reported on their experiences working in companies (see, e.g., the concepts \emph{work/task context}).

\emph{Qualitative evaluation criteria:} To support \emph{credibility} of our findings, we not only inductively built our theory from surveys with 335 software developers, but also deductively included results from related work on expertise and expert performance.
We constantly compared the answers between all three samples and mapped them to overarching concepts and categories.
For the core concepts general/task-specific \emph{knowledge} and \emph{experience}, and the connection of \emph{individual differences}, \emph{work context}, \emph{behavior}, and \emph{performance}, we observed theoretical saturation in the way that those concepts were frequently named and the descriptions did not contradict the relationships we modeled.
However, as we only collected data from three samples of developers, the concepts, and in particular the categories we added in phase 3, have to be validated using more data to achieve a higher level of theoretical saturation. 
In terms of \emph{originality}, we not only contribute a first conceptual theory of SDExp, but also a research design for theory building that other software engineering researchers can adapt and apply.
Regarding the \emph{resonance} of our theory, the feedback, in particular from samples $S2$ and $S3$ with focused questions directly related to theory concepts, was generally positive.
Participants described their participation as a ``very informative experience'' and a ``nice opportunity to reflect''.
However, there was some negative feedback regarding the Java focus, especially in sample $S3$.
Participants were mainly asking why we concentrated on Java, not questioning the general decision to focus on one particular programming language for some questions.
To motivate the \emph{usefulness} of our theory, we refer to Section~\ref{sec:summary-and-future-work}, which contains short summaries of our findings targeting researchers, software developers, and their employers.

\emph{Other limitations:} The qualitative analysis and theory-building was mainly conducted by the first author and was then discussed with the second author.
We tried to mitigate possible biases introduced by us as authors of the theory by embedding our initial GT in related work on expertise and expert performance (see Section~\ref{sec:phase2}) and then again collecting data to further refine the resulting conceptual theory (see Section~\ref{sec:phase3}).
However, when theorizing, there will always be an ``uncodifiable step'' that relies on the imagination of the researcher~\cite{Langley1999, Weick1989}.

\section{Related Work} 
\label{sec:related-work}



Expertise research in \textbf{software engineering} mainly focused on expert recommendation, utilizing information such as change history~\cite{McdonaldAckerman2000, MockusHerbsleb2002, KagdiHammadOthers2008}, usage history~\cite{VivacquaLieberman2000, MaSchulerOthers2009}, bug reports~\cite{AnvikHiewOthers2006}, or interaction data~\cite{FritzMurphyOthers2007, RobbesRothlisberger2013}.
Investigated aspects of software development expertise (SDExp) include programming experience~\cite{SiegmundKaestnerOthers2014}, age~\cite{MorrisonMurphyHill2013}, developer fluency~\cite{ZhouMockus2010}, and desired attributes of software engineers~\cite{LiKoOthers2015} and managers~\cite{KalliamvakouBirdOthers2017}.
Moreover, similar to our study, Graziotin et al. observed that vision and goal-setting are related to developers' performance~\cite{GraziotinWangOthers2015}.
However, as mentioned above, up to now there was no theory combining those individual aspects.

Beside the references mentioned in the description of our theory, the psychological constructs \textbf{personality}, \textbf{motivation}, and \textbf{mental ability} provide many links to theories and instruments from the field of psychology.
To assess developers' \textbf{personality}, e.g., one could employ the \emph{International Personality Item Pool} (IPIP)~\cite{Goldberg1999}, measuring the \emph{big five personality traits}.
There have been many studies investigating the personality of software developers~\cite{CruzdaSilvaOthers2015}.
Cruz et al. conclude in their systematic mapping study that the evidence from analyzed papers is conflicting, especially for the area of individual performance.
Thus, more research is needed to investigate the connection between personality and expert performance.
Our theory can help to identify confounding factors affecting performance, in particular the interplay between an individual's mental abilities, personality, motivation, and his/her general and task-specific knowledge and experience.
The connection between mental abilities, personality, and domain knowledge in expertise development has, for example, been described by Ackerman and Beier's~\cite{AckermanBeier2006}.

The concepts of \textbf{communication} and \textbf{problem-solving skills} have been thoroughly described in psychology literature~\cite{Hargie2006, LockeLathamOthers1990, Robertson2016}.
Researchers can use this knowledge when designing studies about the influence of such skills on the formation of SDExp.
The other two general skills we included in our theory, \textsc{continuous learning} and \textsc{assessing trade-offs}, have also been described by Li et al.~\cite{LiKoOthers2015}, who identified \emph{continuously improving} and \emph{effective decision-making} as critical attributes of great software engineers.

Very closely related to the concept of \textbf{deliberate practice}~\cite{EricssonKrampeOthers1993}, which we included in our theory, is the concept of \emph{self-directed learning}~\cite{MerriamCaffarellaOthers2007} that connects our work to educational research.
Similar to our theory, motivation and self-monitoring are considered to be important aspects of self-directed learning~\cite{MerriamCaffarellaOthers2007}.
To capture the \textbf{motivation} of developers one could adapt ideas from \emph{self-determination theory}~\cite{RyanDeci2000} or McClelland's theory of the \emph{big three motives}~\cite{McClelland1987}.
There also exist instruments like the \emph{Unified Motive Scales (UMS)}~\cite{SchonbrodtGerstenberg2012} to assess human motivation, which can be utilized in studies.
Beecham et al.~\cite{BeechamBaddooOthers2008} conducted a systematic literature review of motivation in software engineering.
While many studies reported that software developers' motivation differs from other groups, the existing models diverge and ``there is no clear understanding of [...] what motivates Software Engineers.''  
Nevertheless, the authors name ``problem solving, working to benefit others and technical challenge'' as important job aspects that motivate developers.
This is very similar to our categories \textsc{work as challenge} and \textsc{helping others}, which we assigned to the concept \textbf{motivation} in our theory.
An area related to motivation is the (perceived) productivity of individual developers \cite{Chrysler1978, MeyerBartonOthers2017} or software development teams \cite{RodriguezSiciliaOthers2012, Gren2017}.
The results from existing studies in this area can be adapted to assess the performance of developers for \textbf{monitoring}, \textbf{feedback}, and \textbf{self-reflection}~\cite{MeyerMurphyOthers2017, TreudeFigueiraFilhoOthers2015}.
Beside their connection to existing software engineering research, those concepts also connect our theory to two additional areas of psychology: \emph{metacognition} (``knowledge about one's own knowledge [... and] performance'')~\cite{FeltovichPrietulaOthers2006} and \emph{self-regulation}~\cite{Zimmerman2006}.

To measure \textbf{mental abilities}, test like the \emph{WAIS-IV}~\cite{WeissSaklofskeOthers2010} or the graphical \emph{mini-q} test \cite{BaudsonPreckel2016} can be employed.
As motivated above, the connection between aging and expertise~\cite{KrampeCharness2006}, and in particular how a (perceived) \emph{age-related performance decline} influences individuals and how they compensate this decline, are important directions for future research.
Considering the phenomenon of global population aging~\cite{LutzSandersonOthers2008}, the number of old software developers is likely to increase in the next decades.
With their experience and knowledge, those developers are a valuable part of software development teams.
However, as our qualitative data suggests, they may become unsatisfied with their jobs and may even drop out of software development. 

To assess the \textbf{performance} of individual software developers, researchers can choose from various existing software metrics~\cite{Jones2008, FentonBieman2015}.
Especially maintainability metrics~\cite{ColemanAshOthers1994} are of interest, because in our study, maintainability was the most frequently named source code property of experts.
Tests about general programming \textbf{knowledge} could be derived from literature about typical programming interview questions~\cite{MonganGigureOthers2013, AzizLeeOthers2012, McDowell2014}.
To assess task-specific Java \textbf{knowledge}, one could rely on commercially available tests like the exams for Oracle's Java certification.
Britto et al.\cite{BrittoSmiteOthers2016} report on their experience measuring learning results and the associated effect on performance in a large-scale software project.
Their results can help measuring the concepts \textbf{education} and \textbf{performance}.

\section{Summary and Future Work}
\label{sec:summary-and-future-work}


In this paper, we presented  a conceptual theory of software development expertise (SDExp).
The theory is grounded in the answers of an online survey with 355 software developers and in existing literature on expertise and expert performance.
Our theory describes various properties of SDExp and factors fostering or hindering its development.
We classified our theory as a \emph{teleological process theory} that views ``development as a repetitive sequence of goal formulation, implementation, evaluation, and modification of goals based on what was learned''~\cite{vandeVenPoole1995}. 
Our task-specific view of SDExp, together with the concept of deliberate practice and the related feedback cycle, fits this framing, assuming that developers' goal is to become experts in certain software development tasks.

We reached a diverse set of experienced and less experienced developers.
However, due to the focus on Java and open source software, future work must investigate the applicability of our results to other developer populations.
We plan to add more results from existing studies in software engineering and psychology to our theory and to conduct own studies based on our theory.
In particular, we want to broaden the scope to include more tasks not directly related to programming.
Nevertheless, the theory is already useful for researchers, software developers, and their employers.
In the following, we will briefly summarize our findings with a focus on those target audiences.

\emph{Researchers:} Researchers can use our methodological findings about (self-assessed) expertise and experience (see Section~\ref{sec:experience-and-expertise}) when designing studies involving self-assessments.
If researchers have a clear understanding what distinguishes novices and experts in their study setting, they should provide this context~\cite{SchwarzOyserman2001} when asking for self-assessed expertise and later report it together with their results.
We motivated why we did not describe expertise development in discrete steps (see Section~\ref{sec:phase2}), but a direction for future work could be to at least develop a standardized description of \emph{novice} and \emph{expert} for certain tasks, which could then be used in semantic differential scales.
To design concrete experiments measuring certain aspects of SDExp, one needs to operationalize the conceptual theory~\cite{HannaySjobergOthers2007}.
We already linked certain concepts to measurement instruments such as UMS (motivation), WAIS-IV (mental abilities), or IPIP (personality).
We also mentioned static analysis tools to measure code quality and simple productivity measures such as commit frequency and number of issues closed. 
This enables researchers to design experiments, but also to re-evaluate results from previous experiments.
There are, e.g., no coherent results about the connection of individual differences and programming performance yet.
One could review studies on developers' motivation~\cite{BeechamBaddooOthers2008} and personality~\cite{CruzdaSilvaOthers2015} in the context of our theory, to derive a research design for analyzing the interplay of individual differences and SDExp.

\emph{Developers:} Software developers can use our results to see which properties are distinctive for experts in their field, and which behaviors may lead to becoming a better software developer.
For example, the concept of deliberate practice, and in particular having challenging goals, a supportive work environment, and getting feedback from peers are important factors.
For ``senior'' developers, our results provide suggestions for being a good mentor.
Mentors should know that they are considered to be an important source for feedback and motivation, and that being patient and being open-minded are desired characteristics.
We also provide first results on the consequences of age-related performance decline, which is an important direction for future work.

\emph{Employers:} Employers can learn what typical reasons for demotivation among their employees are, and how they can build a work environment supporting the self-improvement of their staff.
Beside obvious strategies such as offering training sessions or paying for conference visits, our results suggest that employers should think carefully about how information is shared between their developers and also between the development team and other departments of the company.
Facilitating meetings that explicitly target information exchange and learning new skills should be a priority of every company that cares about the development of their employees.
Finally, employers should make sure to have a good mix of continuity and change in their software development process, because non-challenging work, often caused by tasks becoming routine, is an important demotivating factor for software developers.


\begin{acks}
We thank the survey participants, Bernhard Baltes-G\"otz, Daniel Graziotin, and the anonymous reviewers for their valuable feedback.
\end{acks}


\bibliographystyle{ACM-Reference-Format}
\bibliography{literature}

\end{document}